\documentclass{article}
\usepackage{graphicx}
\usepackage{epsfig}
\usepackage{cite}
\usepackage{color}
\usepackage{amsfonts}
 \usepackage{amssymb}

\date{\today}

\newcommand{\insertplot}[5]{\begin{figure}
 \hfill\hbox to 0.05in{\vbox to #5in{\vfill
 \inputplot{#1}{#4}{#5}}\hfill}
 \hfill\vspace{-.1in}
 \caption{#2}\label{#3}
 \end{figure}}
 \newcommand{\inputplot}[3]{
 \special{ps: plotfile #1}
}
\newcounter{fig}

\newcommand{\ee}{\end{equation}}
\newcommand{\eea}{\end{eqnarray}}
\newcommand{\be}{\begin{equation}}
\newcommand{\bea}{\begin{eqnarray}}

\begin{document}
\title{\bf  Black hole spontaneous scalarisation \\ with a positive cosmological constant } 

\author{
{\large Yves Brihaye}$^{\dagger}$,
{\large Carlos Herdeiro}$^{\ddagger}$ 
{\large } 
and {\large Eugen Radu}$^{\ddagger}$  
\\ 
\\
$^{\dagger}${\small Physique-Math\'ematique, Universite de
Mons-Hainaut, Mons, Belgium}
\\
$^{\ddagger}$ {\small Departamento de Matem\'atica da Universidade de Aveiro and } 
\\ {\small  Centre for Research and Development  in Mathematics and Applications (CIDMA),}   
\\
   {\small Campus de Santiago, 3810-183 Aveiro, Portugal}
}

\date{October 2019}

\maketitle 

\begin{abstract}  
A scalar field non-minimally coupled to certain geometric [or matter] invariants which are sourced by [electro]vacuum black holes (BHs) may spontaneously grow around the latter, due to a tachyonic instability. This process is expected to lead to a new,  dynamically preferred, equilibrium state: a scalarised BH. The most studied geometric [matter] source term for such \textit{spontaneous BH scalarisation}  is the Gauss-Bonnet quadratic curvature [Maxwell invariant]. This phenomenon has been mostly analysed for asymptotically flat spacetimes. Here we consider the impact of a positive cosmological constant, which introduces a cosmological horizon.  The cosmological constant does not change the local conditions on the scalar coupling for a tachyonic instability of the scalar-free BHs to emerge. But it leaves a significant imprint on the possible new scalarised BHs.  It is shown that no scalarised BH solutions exist, under a smoothness assumption, if the scalar field is confined between the BH and cosmological horizons. Admitting the scalar field can extend beyond the cosmological horizon, we construct new scalarised BHs. These are asymptotically de Sitter in the (matter) Einstein-Maxwell-scalar model, with only mild difference with respect to their asymptotically flat counterparts. But in the (geometric) extended-scalar-tensor-Gauss-Bonnet-scalar model, they have necessarily non-standard asymptotics, as the tachyonic instability dominates in the far field. This interpretation is supported by the analysis of a test tachyon on a de Sitter background.
\end{abstract}

\section{Introduction}
 
The ground state of Einstein's gravity with a positive cosmological constant is \textit{de Sitter} (dS) spacetime.
Solutions of Einstein's gravity, or generalisations thereof,  with dS asymptotics are of interest for various reasons.
Firstly and foremost, observational evidence supports that our Universe is undergoing an accelerated expansion~\cite{Perlmutter:1998np,Riess:1998cb}.  The simplest theoretical modelling of such observations consists on assuming a small positive vacuum energy, $i.e.$ a cosmological constant $\Lambda>0$, implying the physical Universe is asymptotically dS.  Secondly, dS spacetime plays a central role in the theory of primordial inflation, the very rapid accelerated expansion in the early Universe, which is now part of the standard cosmological model.
Finally, from a theoretical perspective, the proposal of a holographic duality 
between quantum gravity in dS spacetime and a conformal field theory on the boundary of dS spacetime \cite{Strominger:2001pn,Witten:2001kn} further stimulated the analysis of asymptotically dS spacetimes.
 
Within the classical solutions of gravitating fields in asymptotically  dS spacetimes, the case of black holes (BHs) is especially interesting, as BHs are, in many ways, the gravitational atoms. One may wonder, for instance, how much dS asymptotics may spoil the celebrated simplicity of BHs in electrovacuum general relativity~\cite{Chrusciel:2012jk}, where famously BHs have no hair, in the sense they have no multipolar freedom. As in the asymptotically flat case beyond electrovacuum~\cite{Herdeiro:2015waa}, including additional degrees of freedom and couplings allows a richer landscape of dS BHs. Let us give some examples.

Concerning scalar hair, a number of no-hair results applicable for real scalar fields in asymptotically flat BHs  still hold for $\Lambda>0$~ \cite{Cai:1997ij,Torii:1998ir,Torii:1999uv,Bhattacharya:2007zzb}.
This covers, for instance, models with a
positive semidefinite,
convex scalar potential; or even  non-minimally coupled cases,
provided the scalar
field potential is zero or quadratic \cite{Winstanley:2005fu}.
BHs with scalar hair exist, nonetheless,  
 if the scalar field potential
is non-convex \cite{Torii:1999uv}.
Remarkably, for a  conformally coupled scalar field
with a quartic self-interaction potential there is an exact (closed form)
 hairy BH solution \cite{Martinez:2002ru}.  
As another example, dS  BHs with Skyrme hair have been reported in~\cite{Brihaye:2005an}.
On the flip side, somewhat unexpectedly,  spherically symmetric boson stars, which are self-gravitating, massive, complex scalar fields~\cite{Schunck:2003kk}, do not possess dS generalisations~\cite{Cai:1997ij}, which may prevent the existence of asymptotically dS BHs with synchronised hair~\cite{Herdeiro:2014goa}.
Turning now to the case of vector hair,  dS BHs with Yang-Mills hair have been discussed in~\cite{Torii:1995wv, 
Breitenlohner:2004fp}, while dS BHs with (real) Proca hair are not possible~\cite{Bhattacharya:2007zzb}.
Finally,   sphalerons and  (non-Abelian) magnetic monopoles inside dS BHs are discussed in~\cite{Brihaye:2006kn}. 

The existence of a hairy BH solution does not guarantee \textit{per se} any sort of dynamical viability of such solution, which is, of course, key for the physical relevance of the BH. But a quite generic dynamical mechanism to obtain new hairy BHs that co-exist and are dynamically preferred to the standard General Relativity (GR) electrovacuum BH solutions of Einstein's gravity has been recently under scrutiny: the phenomenon of BH \textit{spontaneous scalarisation}. This phenomenon is induced by non-minimal couplings which allow circumventing well-known no-hair theorems. The non-minimal coupling is typically between a real  scalar field $\phi$ and some source
term ${\cal I}$, which can trigger a repulsive gravitational effect, via an effective tachyonic mass for $\phi$.
As a result, the GR solutions are unstable against scalar perturbations in regions where the source term is significant, dynamically developing scalar hair, $i.e.$ \textit{spontaneously scalarising}.

Various expressions of   ${\cal I}$ have been considered in the literature, that fall roughly into two types:  ${\cal I}$ is a geometric invariant, such as the Gauss-Bonnet invariant~\cite{Silva:2017uqg,Doneva:2017bvd,Antoniou:2017acq,Antoniou:2017hxj,Myung:2018iyq,Blazquez-Salcedo:2018jnn,Doneva:2018rou,Minamitsuji:2018xde,Silva:2018qhn,Brihaye:2018grv,Macedo:2019sem,Doneva:2019vuh,Myung:2019wvb,Andreou:2019ikc,Minamitsuji:2019iwp,Cunha:2019dwb,Konoplya:2019fpy}, the Ricci scalar for non-conformally invariant BHs~\cite{Herdeiro:2019yjy}, or the Chern-Simons invariant~\cite{Brihaye:2018bgc}; or  ${\cal I}$ is a ``matter" invariant, such as the Maxwell $F^2$ term 
\cite{Herdeiro:2018wub, Myung:2018vug,Myung:2018jvi,Fernandes:2019rez,Brihaye:2019kvj,Myung:2019oua,Konoplya:2019goy,Fernandes:2019kmh}. This phenomenon is actually not exclusive of scalar fields~\cite{Ramazanoglu:2019gbz}. It would be therefore interesting to understand the impact of a positive cosmological constant in this phenomenon, and if it can lead, dynamically to hairy BHs in a dS Universe. Work in this direction was reported in~\cite{Bakopoulos:2018nui}.

The goal of this paper is to assess the impact of a positive cosmological constant in the BH spontaneous scalarisation phenomenont, considering the two paradigmatic cases in the literature, but augmented with $\Lambda>0$. We shall then focus on BHs in Einstein-Maxwell-scalar-$\Lambda$ (EMS-$\Lambda$) and extended-Scalar-Tensor-Gauss-Bonnet-scalar-$\Lambda$ (eSTGB-$\Lambda$) models, which, 
for $\Lambda=0$, both allow for BH scalarisation to occur. As we shall see, the impact of the positive cosmological constant is substantially different in the two cases, which is related to the nature of the tachyonic instability, which for the matter model is asymptotically quenched, leading to scalarised asymptotically dS charged BHs, but for the geometric model it is not, leading to a non-asymptotically dS geometry.

This paper is organised as follows. In section~\ref{section2} we discuss the general framework, introduce the two models and the ansatz for the fields, discuss the conditions for scalarisation to occur and scalarised BHs to exist, providing the choice of the non-minimal coupling that shall be used in our work. We also analyse the behaviour of a tachyonic scalar field on dS spacetime that will be relevant for our results. We end this section with a no-go theorem for smooth scalar hair confined between the BH and cosmological horizon. In sections~\ref{section3} and \ref{section4} we describe, respectively, the matter and the geometric model. In each case we start with the construction of the zero modes, the scalar clouds on the scalar-free BH, and then discuss some properties of the non-linear scalarised BH solutions. Section~\ref{section5} provides some final remarks.

\section{The general framework}
\label{section2}

\subsection{Models and ansatz}
The considerations in this work apply to a family of  models described by the following action (setting $c=G=1$):
\begin{eqnarray}
\label{actionS}
\mathcal{S}=- \frac{1}{16 \pi} \int  d^4 x \sqrt{-g} 
\left[
  R-2 \Lambda  
- 2(\nabla \phi)^2 
-  f(\phi) {\cal I}(\psi;g)
\right] \ ,
\end{eqnarray}
where $R$ is the Ricci scalar, $\Lambda  >0$ is the cosmological constant, $\phi$ is  a real scalar field, 
 $f(\phi) $ is the  {\it coupling function}
and 
${\cal I}$ is the {\it source term}. The latter may depend only on the spacetime metric $g_{\mu\nu}$ or also on extra matter fields, collectively denoted by $\psi$.
The corresponding equation of motion for the scalar field  
and the metric tensor read
\begin{eqnarray}
\label{eq-phi}
&&
\Box \phi=f_{,\phi}\frac{{\cal I}}{4}~,
\\
&&
\label{eq-E}
R_{\mu  \nu}-\frac{1}{2}g_{\mu\nu}+\Lambda g_{\mu\nu}=2T_{\mu\nu} \  , \qquad
{\rm where}~~T_{\mu\nu} = T_{\mu\nu}^{(\phi)}+T_{\mu\nu}^{(\psi)} \ .
\end{eqnarray}
Here, $ T_{\mu\nu}^{(\phi)}=\partial_\mu \phi \partial_\nu \phi-\frac{1}{2}g_{\mu \nu}(\nabla \phi)^2$ is the scalar field energy-momentum tensor, whereas  $T_{\mu\nu}^{(\psi)}$ is the energy-momentum tensor associated
with the source term in the action (\ref{actionS}).
These equations must, of course, be supplemented with those describing the dynamics of the matter fields $\psi$, if they are present. 

To be more concrete, we shall focus on two specific models within the family~(\ref{actionS}), corresponding to two different choices of source term ${\cal I}$. These are:
\begin{description} 
 \item[{\bf i)}] {\it a ``matter" source:} ~ ${\cal I}={\mathcal L}_{M}\equiv F_{\mu\nu}F^{\mu\nu} $ \ ,~~
  with~$\psi=A_\mu$~~and~~$F_{\mu \nu}=\partial_\mu A_\nu-\partial_\nu A_\mu$  ~,
\item[{\bf ii)}] {\it a geometric source:}
~~ ${\cal I}={\mathcal L}_{GB}\equiv R^2 - 4 R_{\mu \nu}R^{\mu \nu} 
            + R_{\mu \nu \rho\sigma}R^{\mu \nu \rho\sigma}$ \ .
\end{description}  
We shall refer to these models, respectively, as the Einstein-Maxwell-Scalar-$\Lambda$ (EMS-$\Lambda$) 
model and the extended Scalar-Tensor-Gauss-Bonnet-$\Lambda$ model (eSTGB-$\Lambda$). For the former model, the equations of motion~(\ref{eq-phi})-(\ref{eq-E}) are supplemented by the Maxwell equations for the electromagnetic field 
\begin{eqnarray}
\label{eqM}
\partial_{\mu}(\sqrt{-g}f(\phi)F^{\mu\nu})=0 \ ,
\end{eqnarray}
while the energy-momentum tensor associated to the source term reads
\begin{equation}
T_{\mu\nu}^{(\psi)}=
f(\phi) \left(F_{\mu\rho}F_{\nu}^{~\rho} 
- \frac{1}{4}g_{\mu\nu} F_{\rho\sigma} F^{\rho\sigma}\right) \ .
\end{equation}
For the latter model, no extra matter fields are present ($\psi=0$), and the energy-momentum tensor associated to the source term reads
 \begin{eqnarray}
\label{Teff2}
 T_{\mu\nu}^{(\psi)} =  - 2\alpha P_{\mu\gamma \nu \alpha}\nabla^\alpha \nabla^\gamma f(\phi) \ ,
 \end{eqnarray}  
where
 \begin{eqnarray}
\nonumber
		P_{\alpha\beta\mu\nu}  = -\frac14 \varepsilon_{\alpha\beta\rho\sigma} R^{\rho\sigma\gamma\delta} \varepsilon_{\mu\nu\gamma\delta} 
		 = R_{\alpha\beta\mu\nu}+ g_{\alpha\nu} R_{\beta\mu} - g_{\alpha\mu} R_{\beta\nu} + g_{\beta\mu} R_{\alpha\nu}-g_{\beta\nu} R_{\alpha\mu} 
			+\frac{R}{2} \left( g_{\alpha\mu}g_{\beta\nu} - g_{\alpha\nu}g_{\beta\mu}\right) \ .
 \end{eqnarray}  

In order to find solutions of the model~(\ref{actionS}), whatever its concrete realisation, an appropriate, sufficiently general ansatz must be chosen. In asymptotically dS spacetimes different coordinate systems serve different purposes; we choose static coordinates. The advantage of these (simple) coordinates is their  independence on a certain ``time" coordinate, which is a adapted to the Killing vector field which is timelike in the static patch.  This coordinate system is computationally convenient, since the relevant equations of motion in our problem reduce to ordinary differential equations; it hides, however, the  cosmological expansion and the fact that the spacetime is not stationary. The metric ansatz in static coordinates is of the form
\begin{eqnarray}
\label{metric}
 ds^2=-e^{-2\delta(r)} N(r)  dt^2+\frac{dr^2}{N(r)}+r^2(d\theta^2+\sin^2\theta d\varphi^2) \ ,
\end{eqnarray} 
where a convenient parametrisation of the 
metric function $N(r)$ is
\begin{equation} 
N(r)\equiv 1-\frac{2m(r)}{r}- \frac{\Lambda r^2}{3} \ .
\label{Nr}
\end{equation}
Empty de Sitter spacetime corresponds to $\delta(r)=0$  and $m(r)=0$. It has a cosmological horizon at $r=\sqrt{3/\Lambda}$. The Schwarzschild-de Sitter (SdS) solution, on the other hand, which represents a neutral BH in an accelerating Universe has 
\begin{equation}
\delta(r)=0 \ , \qquad  m(r)=M={\rm constant} \ .
\label{SdSsol}
\end{equation}
In the case of the EMS-$\Lambda$ model, we shall be interested in electrically charged BHs. Then, an ansatz for the electromagnetic 4-potential must be set. We shall restrict ourselves to a purely electric gauge potential, 
\begin{eqnarray}
\label{A}
A=V(r) dt \ .  
\end{eqnarray}
The choices 
\begin{eqnarray}
\label{RN}
\delta(r)=0\ , \qquad 
m(r)=M-\frac{Q^2}{2r}  \ , \qquad 
~~V(r)= \frac{Q}{r} \ ,
\end{eqnarray}
yield the Reissner-Nordstr\"om-de Sitter (RNdS)
BH, where $M$
and $Q$ are the gravitational mass 
and the total electric charge, respectively 
(whose definition is subtle for a dS background 
\cite{Astefanesei:2003gw}).
A discussion of this solution can be found in~\cite{Brill:1993tw,Romans:1991nq}. Finally in all cases we shall consider the scalar  field is a function of $r$ only:
\begin{equation}
\phi=\phi(r) \ .
\label{fi}
\end{equation}

With the ansatz (\ref{metric}), (\ref{A}) and (\ref{fi}) we aim at finding nonsingular, asymptotically dS spacetimes containing a BH. The function $N(r)$ will have (at least) two zeros, corresponding to the BH horizon at $r=r_h>0$ and the cosmological horizon located at $r=r_c>r_h>0$. Both these hypersurfaces are merely coordinate singularities, where all curvature invariants are finite. A nonsingular extension across both of them can be found. Both functions $N(r)$ and $e^{-2\delta(r)}$ are strictly positive between these horizons. We shall also assume that all matter fields (together with their first and second derivatives) are smooth at both BH and cosmological horizons.
Outside the cosmological horizon, $N(r)$ changes sign, such that $r$ becomes a timelike coordinate. To assure standard dS asymptotics, we require $m(r)\to M$ asymptotically outside the cosmological horizon, where the constant $M$ is the BH mass, as can be proven by using the quasilocal formalism and approach in 
\cite{Balasubramanian:2001nb}.\footnote{For this purpose,  the action (\ref{actionS}) is
supplemented with a boundary counterterm, the BH mass being computed outside the horizon, at future/past infinity.} Moreover, we  assume that the metric function $\delta(r)$ vanishes in the far field, decaying faster than $1/r^3$. The matter field(s)  asymptotic behaviour, on the other hand, will result from the field equations and, as we shall see, it will not always be compatible with the assumed standard dS asymptotics.

Both the event  and the cosmological
horizons have their own thermodynamical properties.
For example, the Hawking temperature, $T_H$ and horizon area $A_H$ of each horizon is, 
\begin{eqnarray}
T_H^{(h,c)}=\frac{1}{4\pi} e^{-\delta(r)} |N'(r)| \Big|_{r=r_h,r_c} \ ,
\qquad 
A_H^{(h,c)}=4\pi r^2|_{r=r_h,r_c}\ .
\end{eqnarray}  
Generically $T_H^{(c)} \neq T_H^{(c)}$; thus two horizons are not in thermal equilibrium.

\subsection{Conditions for scalarisation and scalarised BHs; choice of $f(\phi)$}
The mechanism allowing for a dynamical evolution between a scalar-free BH and a scalarised one is, in principle, the same as for the case of asymptotically flat BHs. This has been described in various references, $e.g.$~\cite{Silva:2017uqg,Herdeiro:2018wub}, but we shall briefly spell it out to keep this paper self-contained. 

We assume that the model admits \textit{scalar-free} solutions; that is,  $\phi=0$ is a solution of (\ref{eq-phi}). 
This implies the condition 
\begin{eqnarray}
\label{condx}
\frac{d f}{d \phi}\Big |_{\phi=0}=0 \ .
\end{eqnarray} 
The BH solution with $\phi=0$ is a standard $\Lambda$-electrovacuum solution of Einstein's gravity. For the two models we shall be interested, the scalar-free solution is either the RNdS BH or the SdS BH.

We also assume the model admits scalarised solutions, with ${\phi \neq 0}$. These solutions form a family, that can be labelled by an extra parameter (say, the value of the scalar field at the horizon) that is continuously connected to the scalar-free solution, approaching it as the extra parameter approaches the value for the scalar-free $\Lambda$-electrovacuum solution. One can further impose that the latter solution is unstable against scalar perturbations, such that the scalarised  solution is dynamically preferred. Considering a small-$\phi$   expansion of the coupling function (since one is dealing with a linear analysis in $\phi$)
\begin{eqnarray}
\label{f-phi-small}
f(\phi)=f|_{\phi=0}+ \frac{1}{2} \frac{d^2 f}{d \phi^2}\Big |_{\phi=0}   \phi^2+\mathcal{O}( \phi^3) \ ,
\end{eqnarray}
 the linearised form of (\ref{eq-phi}) reads
\begin{eqnarray}
\label{eq-phi-small}
(\Box-\mu_{\rm eff}^2) \phi =0 \ , \qquad {\rm where}~~ \mu_{\rm eff}^2=   
\frac{1}{4}\frac{d^2 f}{d \phi^2}\Big |_{\phi=0} {\cal I}  \ .
\end{eqnarray}
Thus, the scalar-free solution is unstable if $ \mu_{\rm eff}^2<0$; that is there is a tachyonic instability triggered by a negative {\it effective} mass squared of the scalar field.

Taking into account our specific models, we note that for the RNdS BH, 
\begin{equation}
\mathcal{I}=F_{\mu\nu}F^{\mu\nu}=-\frac{Q^2}{r^4}<0 \ ,
\label{maxi}
\end{equation}
whereas for a SdS BH,
\begin{equation}
\label{gbi}
\mathcal{I}=R^2 - 4 R_{\mu \nu}R^{\mu \nu} 
            + R_{\mu \nu \rho\sigma}R^{a\mu \nu \rho\sigma}= \frac{48 M^2}{r^6}+\frac{8}{3}\Lambda^2>0 \ .
\end{equation}
Now we need a specific choice of the coupling function $f(\phi)$. We shall focus on a quadratic  coupling function, the simplest function that contains the necessary term in~(\ref{f-phi-small}):
\begin{eqnarray}
\label{cf}
 f(\phi)=a_0-\alpha \phi^2 \ .
\end{eqnarray}
The first constant is taken as  $a_0=1$ for the EMS-$\Lambda$ model and an arbitrary value for the eSTGB-$\Lambda$ case.
The second constant,  $\alpha$, defines the sign of $d^2f(\phi)/d\phi^2$, and hence that of $\mu^2_{\rm eff}$. In fact,  $\mu_{\rm eff}^2=-\alpha {\cal I}/2$. From (\ref{eq-phi-small})-(\ref{cf}), the existence of a tachyonic instability requires
\begin{equation}
\alpha<0 \ \ \ {\rm for \ EMS-}\Lambda  \qquad {\rm and} \qquad \alpha>0 \ \ \  {\rm for \ eSTGB-}\Lambda \ .
\end{equation}
Observe that $\alpha$ is dimensionless for the EMS-$\Lambda$ model and has dimension
${\rm [length]}^2$ for the eSTGB-$\Lambda$ model.\footnote{In this work we shall plot various quantities
which are invariant under a scaling of the radial coordinate $r\to \lambda r$ (with $\lambda>0$),
and for the eSTGB-$\Lambda$ model, also $\alpha\to \alpha/\lambda^2$ (and various global quantities scaling accordingly).}

Solving (\ref{eq-phi-small}) on the $\Lambda$-electrovacuum BH spacetimes and the above coupling function is an eigenvalue problem.  The solutions that obey the appropriate boundary conditions describe zero modes or {\it scalar clouds}. 
For each choice of   ${\cal I}$,  
they exist 
 for a specific (discrete) set of global charges. These linear zero modes mark
the onset of the instability triggered by the scalar field perturbation and the branching off towards a new family of fully non-linear solutions describing scalarised BHs.

Ensuring the above instability of the scalar-free solutions can one really guarantee the existence of a new set of scalarised solutions? Although this can only be done by explicitly computing the latter,  some Bekenstein-type identities put constraints on the models that can have scalarised solutions. Let us provide three examples.

As a first example, we integrate eq.~(\ref{eq-phi})
along a hypersurface $V$ bounded by
the BH horizon and the cosmological horizon. 
Since the contribution of the  boundary terms vanishes for smooth configurations, 
this results in  the identity
\begin{eqnarray}
\label{eq1}
 \int_V d^4 x \sqrt{-g} 
f_{,\phi }  {\cal I}
=0~.
\end{eqnarray}
Assuming that the source term ${\cal I}$
does not change the sign between the BH and cosmological horizons, which is true in the test field limit for the specific models described above, this identity implies that  $f_{,\phi}$, which equals $-2\alpha \phi$ for choice~(\ref{cf}), has to change sign in the interval $r_h<r<r_c$ for non-trivial scalar fields to be possible. Thus, the the number of nodes $k\in \mathbb{N}_0$ of the scalar field in between the two horizons must be $k\geqslant 1$. In this work, for simplicity, we shall focus on solutions with the minimal number of nodes, $k=1$.

As a second example, we multiply
eq.~(\ref{eq-phi})
by $f_{,\phi}$. 
After integrating by parts and using the divergence theorem, 
this results is 
\begin{eqnarray}
\label{eq2}
 \int_V d^4 x \sqrt{-g} 
\left(
f_{,\phi \phi} (\nabla  \phi)^2+\frac{1}{4}f_{,\phi}^2
{\cal I}
\right)=0~.
\end{eqnarray}
Again, if the source term ${\cal I}$
does not change the sign between the BH and cosmological horizon
 this identity requires $f_{,\phi \phi}$ and $\mathcal{I}$ to have the opposite sign in some interval between the two horizons, for a non-trivial scalar field profile to exist. From (\ref{eq-phi-small}) and for our coupling this is precisely the requirement that $\mu^2_{\rm eff}$ is negative. Thus, a non-tachyonic scalar field with $\mu^2_{\rm eff}>0$ everywhere cannot yield scalar hair 
(at least as a test field on the standard $\Lambda$-electrovacuum BHs).

A third, related, example is found by multiplying (\ref{eq-phi})
by $\phi$, the integration resulting in 
\begin{eqnarray}
\label{eq3}
 \int_V d^4 x \sqrt{-g} 
\left(
  (\nabla  \phi)^2+\frac{1}{4} \phi f_{,\phi}
{\cal I}
\right)=0 \ .
\end{eqnarray}
Similarly, this now implies that 
$\phi f_{,\phi}$  and $\mathcal{I}$ must have the opposite sign somewhere in  the interval $r_h<r<r_c$. For our coupling this leads to the same conclusion as the identity~(\ref{eq2}).

\subsection{A  tachyon on dS spacetime}
\label{tachyonsec}
From the above discussion, a scalar field must have a tachyonic behaviour somewhere in between the BH and cosmological horizon, for scalar hair to exist. What is the asymptotic behaviour, beyond the cosmological horizon, of such a tachyon? This question, which impacts on our findings of the next sections, can be tackled by considering the massive Klein-Gordon equation, $(\Box-\mu^2 ) \phi =0$, with $\mu^2 =$constant, as a test field on an empty de Sitter spacetime. A closed form solution can be found, which	consists of the sum  of two modes:
 \begin{eqnarray}
\label{sol-dS}
 \phi(r)= \frac{1}{r}P_u \left( \frac{r}{r_c} \right) 
+ 
\frac{s}{r}Q_u \left( \frac{r}{r_c} \right) \ , \qquad {\rm where}~~
u\equiv \frac{3\chi-1}{2} \qquad {\rm and} \ \ \  \chi\equiv \sqrt{1- \frac{4\mu^2}{3\Lambda} } \ .
\end{eqnarray}
Here, $P_u$, $Q_u$ are Legendre functions and $s$ is an arbitrary constant.
Both terms in the above solution diverge at $r=0$; $Q_u (r/r_c)$ also diverges at the cosmological horizon, located at $r=r_c= \sqrt{ {3}/{\Lambda}}$.
Thus, in what follows we take $s=0$. Then, the solution in the neighbourhood of the cosmological horizon expands as 
\begin{equation}
\phi(r)=\frac{1}{r_c}-\frac{\mu^2}{2}(r-r_c)+\mathcal{O}(r-r_c)^2 \ .
\end{equation} 
For $r\gg r_c$, on the other hand, the approximate form of $ \phi(r)$ is
 \begin{eqnarray}
\label{sol-dS2}
 \phi(r)\simeq c_+ r^{-\frac{3}{2} \left(1+\chi \right)}
 +  c_{-}r^{-\frac{3}{2} \left(1-\chi\right)} \ , \qquad {\rm where}  \ \  c_{\pm}\equiv \frac{ r_c^{\frac{1 \pm 3\chi}{2}} 
 \Gamma\left(\mp \frac{3\chi}{2} \right)}{
\sqrt{\pi}2^{\frac{1\pm 3\chi}{2}}
 \Gamma \left(\frac{1\mp 3\chi}{2} \right)} \ .
\end{eqnarray}
For a tachyonic field $\mu^2<0$ and $\chi>1$; thus $ \phi(r)$ diverges as $r\to \infty$. Let us stress this conclusion: a tachyonic test field  (solely depending on $r$) that is regular at the cosmological horizon is necessarily asymptotically divergent, and the test field approximation breaks down.

In the presence of a BH, one may expect this asymptotic behaviour to remain, again if one assumes regularity at the cosmological horizon, if the scalar field has an effective tachyonic mass, asymptotically.  This is corroborated by the numerical results in the next Sections. Although in our models $\mu^2_{\rm eff}$ is a function of $r$, the existence (or absence) of an asymptotic tachyonic behaviour 
in the  region  $r\gg r_c$ will source a deviation from standard de Sitter asymptotics.  The $r=0$ singularity of~(\ref{sol-dS}), on the other hand, becomes irrelevant in the presence of a BH horizon.

\subsection{No smooth scalar hair confined within the cosmological horizon}
We have seen that, on the one hand, a tachyonic behaviour is required for the scalar field to be non-trivial in between the BH and the cosmological horizon; on the other hand, an asymptotic tachyonic behaviour will potentially lead to divergences. One may ask, thus, if one could confine the non-trivial scalar entirely within the BH and cosmological horizon, thus excising the potential pathological behaviour. 

If such confined scalar field is smooth, not only it vanishes at the cosmological horizon, but its derivatives, and in particular the first derivative, also vanish therein. Then, one can show that 
for a large class of models, $\phi(r_c)=0=\phi'(r_c)$ imply that $\phi \equiv 0$
 for the whole region $r_h<r<r_c$. The proof goes as follows.
For a scalar field with $k$ nodes in $r_h<r<r_c$, the assumption
$\phi(r_c)=0$ implies the existence of (at least) $k$ local extrema of its profile. Recall $k\geqslant 1$.
Let $r_0$ be the largest root of the equation $\phi'(r)=0$ ($r_0<r_c)$.
Then, integrating the scalar field equation (\ref{eq-phi}) 
between $r_0$ and $r_c$ yields
\begin{eqnarray}
e^{-\delta}N r^2 \phi'\big |_{r_0}^{r_c} =
\frac{1}{4} \int_{r_0}^{r_c} dr~ e^{-\delta}  r^2  f_{,\phi}{\cal I}~.
\label{nch}
\end{eqnarray}
The left hand side of~(\ref{nch}) vanishes. Indeed, a smooth configuration has $N(r_c) e^{-\delta(r_c)} \phi'(r_c) \to 0$; moreover, both $N$ and $e^{-\delta}$ are finite at $r_0$, where $\phi'(r_0)=0$.
However,  for the EMS-$\Lambda$ model and also for the test field limit 
 of the eSTGB-$\Lambda$ model, the integrand of the right hand side does not change the sign in that $r$-interval.
We conclude that $\phi \equiv 0$ for the considered $r-$range. The argument can easily be extended for all interval $r_h<r<r_c$, yielding the advertised result.

\section{The scalarised EMS-$\Lambda$ black holes}
\label{section3}

 \subsection{The zero modes}
 \label{emszero}
For the EMS-$\Lambda$ model, the scalar-free solution is the RNdS BH, given by~(\ref{metric}), (\ref{Nr}) and~(\ref{A}) with~(\ref{RN}) and $\phi=0$. Let us first consider the zero modes of the scalar field perturbations. In this paper we only consider spherical modes.

The small-$\phi$ limit of the scalar field equation (\ref{eq-phi}) on a fixed RNdS background gives
\begin{eqnarray}
\label{p2}
 (r^2 N  \phi')'-
 \frac{\alpha Q^2} {r^2}  \phi=0 \ .
\end{eqnarray} 
For $\Lambda=0$, (\ref{p2}) admits an exact, closed form solution in term of a Legendre function~\cite{Herdeiro:2018wub}
\begin{eqnarray}
\label{ex1} 
\phi(r)=P_u 
\left[
1+\frac{2Q^2(r-r_h)}{r(r_h^2-Q^2)}
\right] \ , \qquad {\rm where} \ \ \ u\equiv \frac{\sqrt{4\alpha+1}-1}{2} \ .
\end{eqnarray}
The leading behaviour of this solution as the asymptotically flat region is approached is
\begin{equation}
\label{ex2}
  \phi (r  \to  \infty)=
{}_2F_1 
\left[
\frac{1-\sqrt{4\alpha+1}}{2},\frac{1+\sqrt{4\alpha+1}}{2},1; \frac{Q^2}{Q^2-r_h^2}
\right]+\mathcal{O}\left(\frac{1}{r}\right) \ .
\end{equation}
Allowing a generic value of $\phi (r  \to  \infty)$, there is a continuum of zero mode solutions, as long as~\cite{Herdeiro:2018wub}
\begin{equation}
\alpha<\alpha_{\rm max}\equiv -\frac{1}{4} \ .
\end{equation} 
The asymptotic value of $ \phi$ is fixed by the ratio $Q/M$.
Requiring, for a given $\alpha$, that the scalar field vanishes asymptotically (i.e. $\phi (r  \to  \infty)=0$),
only a discrete set of values of $Q/M$ is allowed, corresponding to solutions with different node number.

No exact solution of~(\ref{p2}) appears to exist for $\Lambda \neq 0$. 
In the neighbourhood of the BH horizon, however, an approximate (regular) solution can be expressed as a power series in $(r-r_h)$, as 
\begin{eqnarray}
\label{ss1}
 \phi(r) =\phi_h+
\frac{\alpha Q^2 r_h (r_c^2+r_c r_h+r_h^2)\phi_h}{(r_c-r_h)r_h[-r_c r_h^2(r_c+2r_h)
+Q^2(r_c^2+2r_c r_h+3r_h^2)]}(r-r_h)+\mathcal{O}(r-r_h)^2 \ , 
\end{eqnarray}
where $\phi_h$ is the value of the scalar field at the BH horizon, a free parameter.
A similar expression holds in the neighbourhood of the cosmological horizon, with $r_h$ and $r_c$ interchanged and $\phi_h$ replaced by the value of the scalar field at the cosmological horizon, $\phi_c$.\footnote{In the numerics we have set $\phi_h=1$ 
without any loss of generality.}  
 
Performing a numerical integration in the region between the BH and cosmological horizons, our numerical results indicate that for a given RNdS background, as specified $e.g.$ by the dimensionless 
ratios ($Q/M$, $r_c/r_h$), solutions which are regular at both horizons 
exist for a discrete set of  $\alpha$, being labelled by the node number $k>0$. 
Using these solutions, the boundary data at the cosmological horizon is fixed;
we then integrate from the horizon outwards, extending the
solutions to the asymptotic region $r\to \infty$.
For large $r$, an approximate form solution can be found as a power series in $1/r$, with the leading order terms being
\begin{eqnarray}
\label{ss1n}
 \phi(r) = \phi_\infty+\frac{\phi_3}{r^{3}}+ \dots \ ,
\end{eqnarray}
where $\phi_\infty$ and $\phi_3$ are constants fixed by the numerics.

\begin{figure}[t!]
\hbox to\linewidth{\hss%
	\resizebox{9cm}{7cm}{\includegraphics{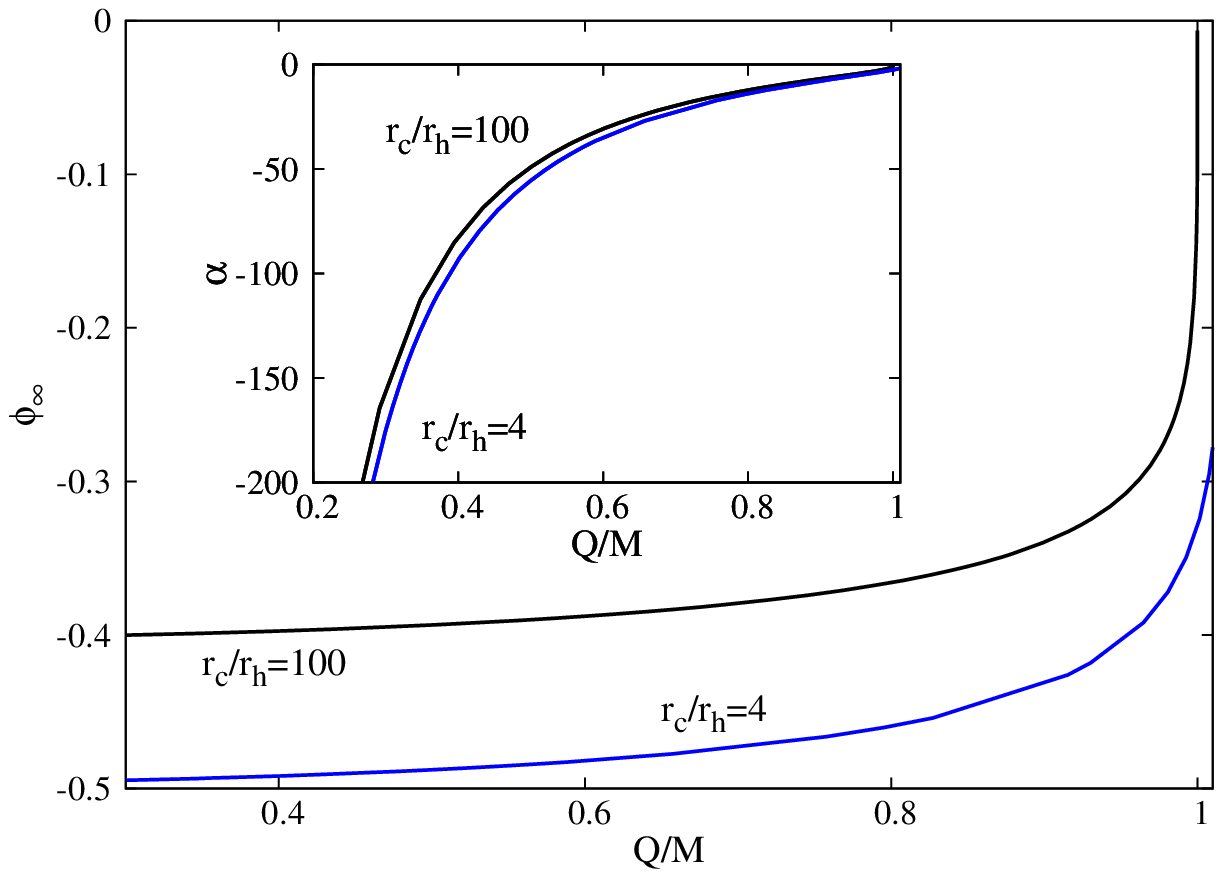}} 
		  \resizebox{9cm}{7cm}{\includegraphics{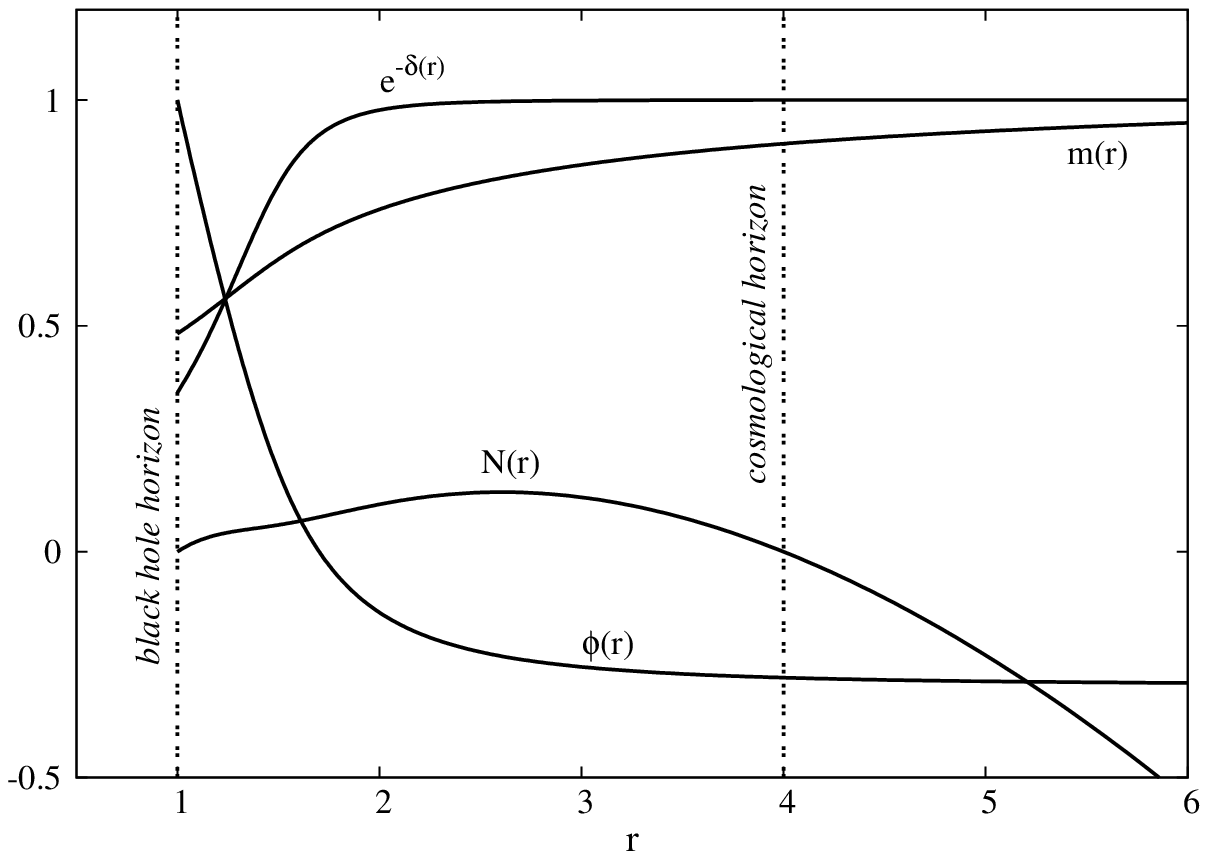}}  
\hss}
\caption{\small 
(Left panel)
Asymptotic value of the scalar field ($\phi_\infty$) (main plot) and critical value of $\alpha$ (inset)
$vs.$ the charge to mass ratio for dS scalar clouds on the  RNdS background, for two illustrative values of $r_c/r_h$. (Right  panel) Radial profiles for the metric functions and electrostatic potential of a typical EMS-$\Lambda$ BH with
$\Lambda>0$.
}
\label{probe-EMs}
\end{figure}

An outstanding fact is that, differently from the $\Lambda=0$ case, solutions with 
$ \phi_\infty \neq 0$ were not found. That is, the scalar field does not vanish asymptotically.  This numerical finding agrees with the analysis in section~\ref{tachyonsec}. Indeed, for the Maxwell case, the effective tachyonic mass vanishes in the far field region, $cf.$~(\ref{maxi}),  and thus (\ref{sol-dS2})
reduces to (\ref{ss1n}).  The behaviour of $\phi_\infty$, as well as the variation of the critical value of $\alpha$ as the BH charge to mass ratio $Q/M$ is varied, is illustrated in Fig.~\ref{probe-EMs} (left panel) for two values of the ratio $r_c/r_h$.
 
 {\small \hspace*{3.cm}{\it  } }

 \subsection{The non-linear solutions}
Let us now consider the non-linear solutions that bifurcate from the RNdS family at the scalar clouds.  The ansatz~(\ref{metric}), (\ref{Nr}),~(\ref{A}) and (\ref{fi}) yields the following set of coupled ordinary differential equations:\footnote{There is also an extra equation, which is a constraint, and can be derived from 
(\ref{ec-m})-(\ref{ec-V}). }
\begin{eqnarray}
\label{ec-m}
&&
m'=\frac{r^2N \phi'^2}{2}  +\frac{e^{2\delta}r^2 V'^2}{2f(\phi)}\ , \qquad \delta'+r \phi'^2=0 \ , 
\\
\label{ec-V}
&&
( f(\phi) e^{\delta}r^2 V')'=0 \ , \qquad 
 (e^{-\delta} r^2 N\phi')'=\frac{e^{\delta}r^2}{2}  \frac{df(\phi)}{d\phi} V'^2 \ .
\end{eqnarray}
The electric potential  can be eliminated from the above equations noticing the existence of a first integral,
\begin{eqnarray}
\label{first-int}
V'= e^{-\delta } \frac{Q}{r^2 f(\phi)}~, 
\end{eqnarray}
where $Q$ is an integration constant interpreted as the electric charge.

The system of equation (\ref{ec-m})-(\ref{ec-V}) will be solved numerically. To do so, we first find  the approximate form of the solutions at the boundary of the domain of integration. Firstly, close to the BH horizon, the relevant functions are approximated as:
\begin{eqnarray}
\label{horizon1}
&&
m(r)=\frac{r_h}{2}-\frac{\Lambda r_h^2}{6}+m_1(r-r_h)+\dots \ , \qquad 
\delta(r)=\delta_h + \delta_1 (r-r_h)+\dots \ ,
\\
&&
\nonumber
\phi(r)=\phi_h + \phi_1 (r-r_h)+\dots\ , \qquad 
V(r)=V_h+v_1 (r-r_h)+\dots \ .
\end{eqnarray}
These expressions depend on the following set of constants: $r_h,\Lambda,m_1,\delta_h,\delta_1,\phi_h,\phi_1,V_h,v_1$. The field equations relate these parameters. We obtain:
\begin{eqnarray}
\label{horizon2}
m_1=\frac{ Q^2}{2r_h^2(1-\alpha \phi_h^2)}\ , \qquad v_1=-\frac{e^{-\delta_h}Q}{(1-\alpha \phi_h^2)r_h^2}\ , \qquad  \phi_1= \frac{\alpha \phi_h e^{2\delta_0}r_h  v_1^2}{1-2m_1-\Lambda r_h^2}\ , \qquad 
\delta_1=-\phi_1^2 r_h \ .
\end{eqnarray}
Thus, the independent parameters are $r_h,\Lambda,\phi_h,\delta_h,V_h$, 
which determine all others.
A similar expression holds at the cosmological horizon which is located at $r=r_c>r_h$, introducing the new independent parameters $\phi(r_c), \delta(r_c),V(r_c)$.
Also, one finds
the following asymptotics of the solutions in the far field:
\begin{eqnarray}
\label{inf1}
&&
m(r)=M-\frac{Q^2 }{2r(1-\alpha^2\phi_\infty^2)}+ \dots\ , \qquad 
\delta(r)=\frac{3 q_s^2}{2r^6}+\dots \ , 
\\
&&
\nonumber
\phi(r)=\phi_\infty+\frac{q_s}{r^3}+\dots \ , \qquad
V(r)=V_\infty+\frac{Q}{(1-\alpha \phi_\infty^2)r}+\dots \ . 
\end{eqnarray}
which introduces the new independent parameters $M,Q,V_0,q_s,\phi_\infty$.\footnote{The value of one of the parameters
$V_h,~V(r_c),~V_\infty$ can be fixed via a gauge transformation.
} 

The field equations for this model (and also the model in the next section) have been solved by the Newton-Raphson method, with
an adaptive mesh selection procedure, with the solver described in~\cite{COLSYS}.
The solutions are found in two steps: first, by integrating from $r_h$ to $r_c$, 
and then from the cosmological horizon to infinity (the region inside the BH horizon is not considered, although it could be studied following~\cite{Brihaye:2016vkv}).
In our approach, both $r_h$ and $r_c$ are input parameter, the corresponding value of $\Lambda$
  resulting from the numerical output. In the following, we shall exhibit some illustrative solutions, which reflect the most relevant properties of the domain of existence studied.

The profile of a typical scalarised RNdS  BH is shown in Fig.~\ref{probe-EMs} (right panel). One checks that $N(r)$ vanishes both at the BH and cosmological horizons; the scalar field starts at a positive value at the BH horizon and is negative at the cosmological horizon, possessing precisely one node; moreover it does not approach zero asymptotically. One also observes that both the mass function $m(r)$  (which is monotonically increasing) and the metric function $e^{-\delta(r)}$ appear to converge for large $r$ suggesting a smooth solution is asymptotically attained.

\begin{figure}[t!]
\hbox to\linewidth{\hss%
	\resizebox{9cm}{7cm}{\includegraphics{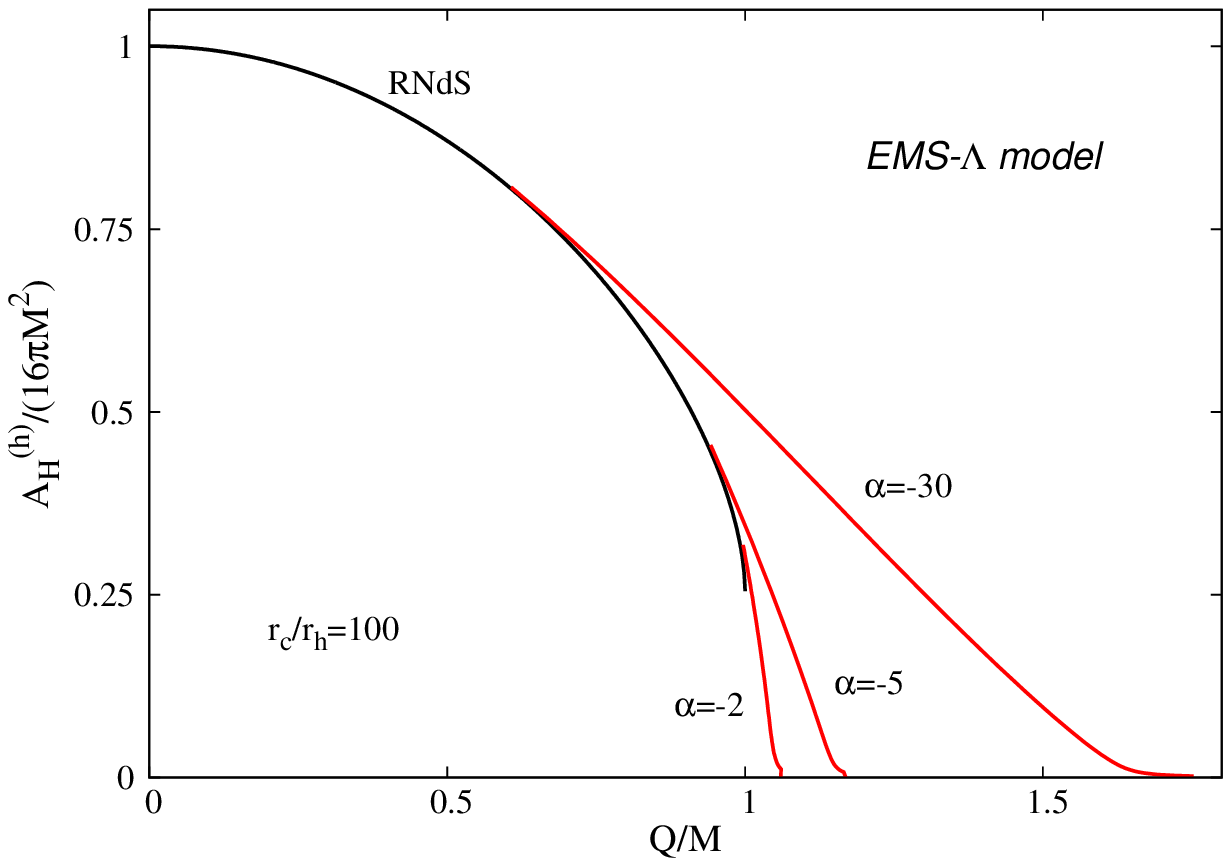}} 
		\resizebox{9cm}{7cm}{\includegraphics{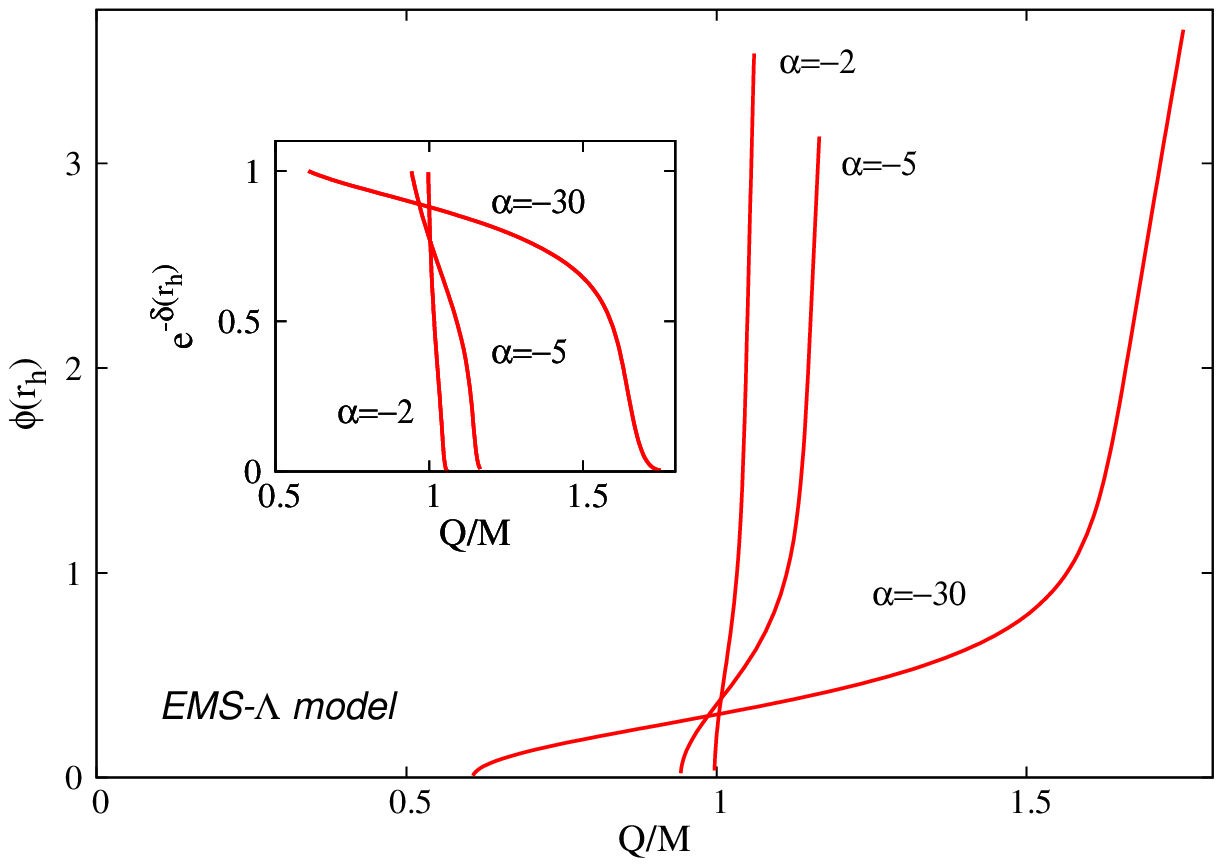}} 
\hss}
\caption{\small 
Normalise horizon area (left panel) and scalar field value at the horizon (right panel) $vs.$ the charge to mass ratio for scalarised EMS-$\Lambda$ BHs, for a fixed value of $r_c/r_h$ and different values of the coupling constant $\alpha$. The right panel also shows the value of the metric function $e^{-\delta{r}}$ at the horizon.}
\label{EMs}
\end{figure}

 Considering now a more global perspective on the full set of computed solutions, 
the emerging picture has some similarities with that found for the $\Lambda=0$ EMS model~
\cite{Herdeiro:2018wub,Fernandes:2019rez},
and can be summarised as follows - see Fig.~\ref{EMs}.
For each $\alpha<\alpha_{\rm max}$, a branch of fully non-linear solutions bifurcates
from a RNdS BH with a particular charge to mass ratio $Q/M$ (and a given ratio $r_c/r_h$). The left panel of Fig.~\ref{EMs} exhibits this bifurcation in a BH (normalised) horizon area diagram $vs.$ the charge to mass ratio. 
One can appreciate that, for a fixed value of $Q/M$, the scalarised solution has a larger BH horizon 
than the corresponding scalar-free solution. Also, 
overcharged solutions exist, just as in the $\Lambda=0$ model. 
Each branch of the scalarised BHs can be specified by the value of the scalar field at the horizon - Fig.~\ref{EMs} (right panel). Each such branch ends at a critical, (likely) singular, configuration: the numerics indicate
the Kretschmann scalar and the horizon temperature diverge,
the BH horizon area vanishes (with $A_H^{(c)}$ still finite),  whereas the mass parameter $M$ stays finite. All these features resemble the $\Lambda=0$ case.

Contrasting with the $\Lambda=0$ case,
the scalarised BHs do not approach precisely the scalar-free solution as $r\to \infty$. Indeed, the scalar field does not vanish as $r\to \infty$,
approaching a constant nonzero value,
a feature anticipated from the analysis of the zero modes.\footnote{Despite this fact, using the approach in \cite{Astefanesei:2003gw}, it can be shown  that
the constant $Q$ can still be  identified with the total electric charge,
as evaluated at future/past infinity.}

 {\small \hspace*{3.cm}{\it  } }

\section{The scalarised eSTGB-$\Lambda$ black holes }
\label{section4}

\subsection{ The zero modes}
 For the eSTGB-$\Lambda$ model, the scalar-free solution is the SdS BH, given by~(\ref{metric}) and (\ref{Nr}) with~(\ref{SdSsol}) and $\phi=0$. Increasing the value of $M$ in de SdS solution implies that  the cosmological horizon (located at the largest root of the equation $N(r)=0$)
shrinks in size, pulled inwards by the gravitational attraction of the BH. 
As a result there is a largest BH, the Nariai solution~\cite{Nariai}, which occurs when
$
M= 1/(3 \sqrt{\Lambda}).
$ 
Spaces with larger values of $M  $ are unphysical,  containing naked singularities.
 Let us again first consider the zero modes of the scalar field perturbations.

Restricting to the small-field limit, equation (\ref{eq-phi-small}) on the SdS background becomes
\begin{eqnarray}
\label{aqw1}
(r^2 N(r) \phi')'+\alpha\frac{6( 2r^6+r_c^2r_h^2(r_c+r_h)^2)}{r^4(r_c^2+r_c r_h+r_h^2)^2}  \phi=0 \ ,
\end{eqnarray}
 where we have eliminated the parameters $M, \Lambda$ in favor of the two horizons radii $r_h,r_c$.
The approximate expression of a regular solution near the BH horizon reads
\begin{eqnarray}
\label{sph1}
\phi(r)=\phi_h + \frac{6 \alpha \phi_h(2r_h^4+r_c^2 r_h^2+r_c^4+2 r_h r_c^3) }{ r_h^3(2r_h^4+r_c r_h^3-r_c^4-2 r_c^3 r_h)}(r-r_h)+\dots  , 
\end{eqnarray}
where $\phi_h$ is the arbitrary constant corresponding to the scalar field value at the horizon.  
A similar expansion exists near the cosmological horizon, which introduces another constant $\phi(r_c)$, instead of $\phi_h$.

\begin{figure}[t!]
\hbox to\linewidth{\hss%
	\resizebox{9cm}{7cm}{\includegraphics{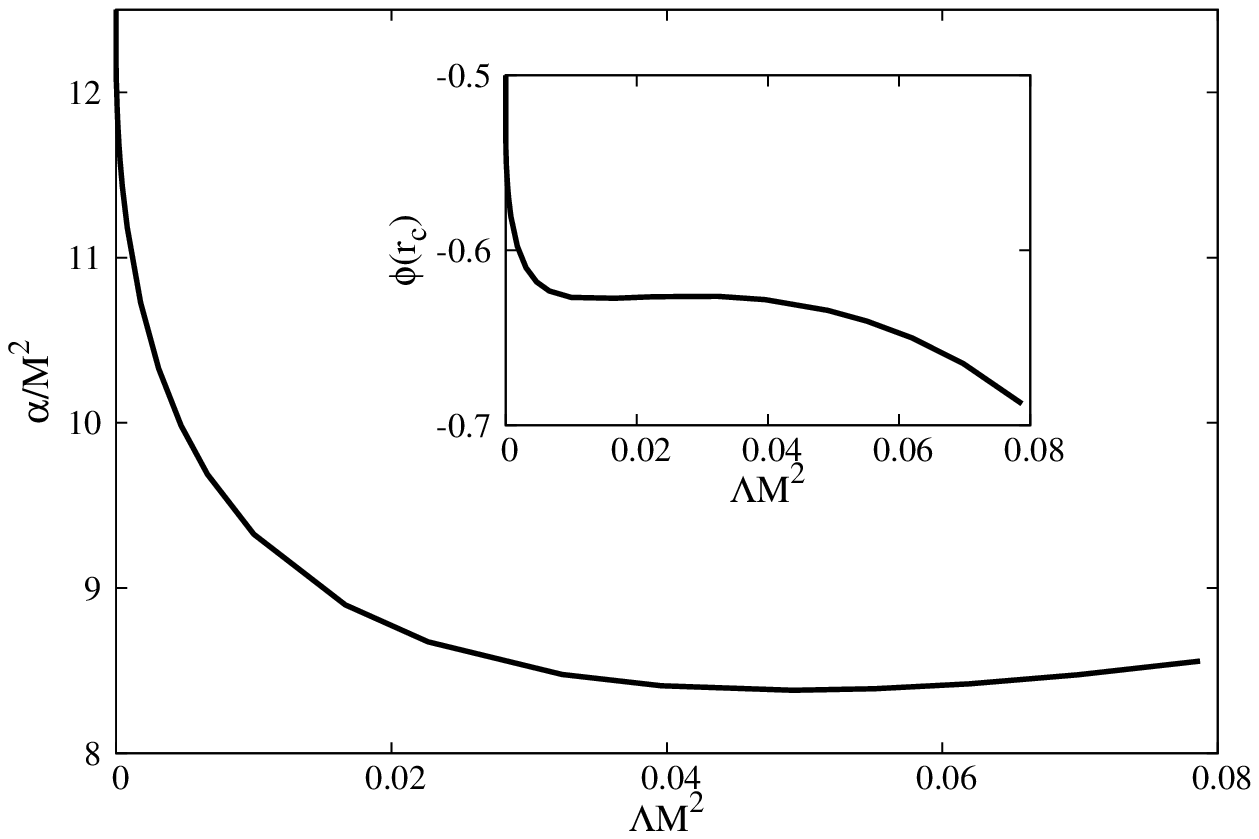}} 
		\resizebox{9cm}{7cm}{\includegraphics{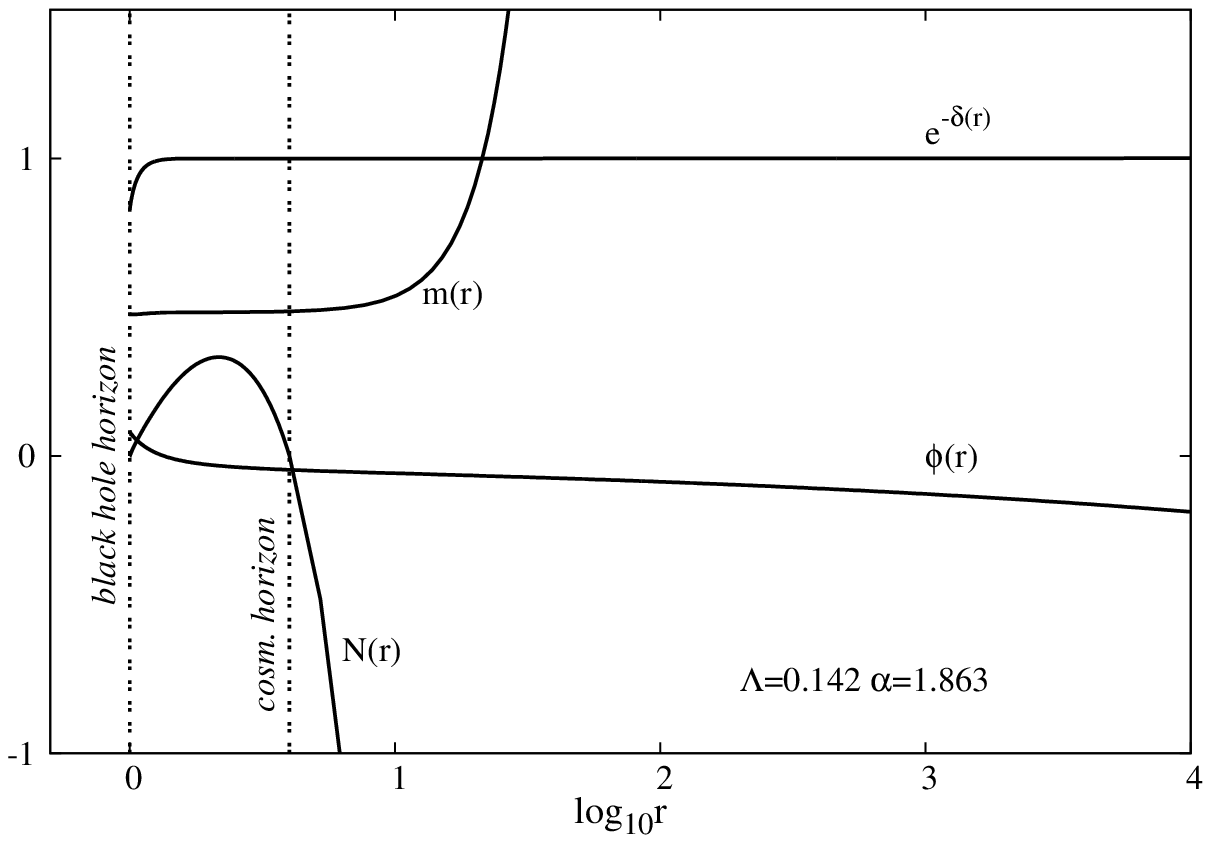}} 
\hss}
\caption{\small 
(Left panel)
The ratio $\alpha/M^2$ $vs.$ the normalised cosmological constant  for the critical SdS BH that supports a spherical cloud with $k=1$ in the eSTGB-$\Lambda$ model. The inset shows the value of the scalar field at the cosmological horizon.  
(Right panel) Radial profile functions for a typical solution of the EGBs-$\Lambda$ model.
}
\label{EGBs-probe}
\end{figure}

Similarly to the case in section~\ref{emszero}, solving the perturbation equation (\ref{aqw1}) 
can be viewed as an eigenvalue problem: 
imposing smoothness for the 
scalar field at  the BH  horizon ($r=r_h$)  and at the cosmological horizon ($r=r_c$)
 selects a discrete set of
 background configurations, specified by the dimensionless ratio $\alpha/M^2$.
For each value of this ratio, a discrete set of scalar profiles is found, labelled by the number of nodes $k>0$. For $\Lambda=0$ these are discussed in~\cite{Silva:2017uqg,Cunha:2019dwb,Hod:2019pmb}.
The dimensionless ratio $\alpha/M^2$ and the scalar field value at the cosmological horizon are shown against the cosmological constant for $k=1$ scalar clouds in Fig.~\ref{EGBs-probe} (left panel). We remark that as $\Lambda\rightarrow 0$,
 the ratio $\alpha/M^2$ does not match the threshold value for the fundamental mode in~\cite{Silva:2017uqg,Cunha:2019dwb,Hod:2019pmb}, which has $k=0$, but rather the first excited state, which has $k=1$. 
 
 {\small \hspace*{3.cm}{\it  } }

There is, however, a key difference between the scalar clouds in this model and those in both the scalar clouds in the asymptotically flat eSTGB model and the EMS-$\Lambda$ model discussed in the previous section.  The scalar clouds always diverge as $r\to \infty$.
That it, for large $r$, the leading terms of the asymptotic  solution of the  eq. (\ref{aqw1})
consist in the sum of two modes
\begin{eqnarray}
 \phi(r)=c_1 r^{-\frac{3}{2}(1+\sqrt{1+ 16 \alpha \Lambda/9})} 
        +c_2 r^{-\frac{3}{2}(1-\sqrt{1+ 16 \alpha \Lambda/9})} + \dots~,
\end{eqnarray}
where $c_1$ and $c_2$ are two constants resulting from the numerics.
The solutions with $c_2= 0$ would possess the right asymptotic behaviour; but these do not arise when integrating from the near BH region. This behaviour is interpreted from the discussion in section~\ref{tachyonsec}.
Since, from~(\ref{gbi}),  $\mu^2_{\rm eff}\stackrel{r\rightarrow \infty}{ \to} -8\alpha/ \Lambda<0 $ in the eSTGB-$\Lambda$ model, eq. (\ref{sol-dS2}) implies that the scalar field necessarily diverges asymptotically.
While the BH horizon indeed `cures' the singularity inside the  cosmological horizon,
no solutions with $\mu^2_{\rm eff}<0$ exist which are regular at both horizons $and$ for large $r$. Thus, the discussion of zero modes already anticipates that BH scalarisation in the eSTGB-$\Lambda$ model will change the de Sitter asymptotics. Moreover, the test field approximation breaks down outside the cosmological horizon.

\subsection{ Including backreaction}
With the ansatz (\ref{metric}) and (\ref{fi}),
a suitable combination of the equations of motion leads to 
first  order equations for the metric functions,  
$m'=F_1(N, \phi,\phi')$,
$\delta'=F_2(N, \phi,\phi')$
and a second order equation for the scalar field,
$\phi''=F_3(N, \phi,\phi')$. These are the equations used  in our numerical approach,  but the expression for the $F_i$ are long and unenlightening; we shall therefore not include them here. 

As for the EMS-$\Lambda$ model, the eSTGB-$\Lambda$ model possesses BH solutions with a non-trivial scalar field which are interpreted as the non-linear realisations of the zero modes discussed above. The profile of a typical solutions is shown in Fig.~\ref{EGBs-probe} (right panel). Comparing with the corresponding profiles for the EMS-$\Lambda$ case, displayed in Fig.~\ref{probe-EMs} (right panel) one both observes similarities and differences. Again, $N(r)$ vanishes both at the BH and cosmological horizons; the scalar field starts again at a positive value at the BH horizon and is negative at the cosmological horizon, possessing precisely one node. Again, it does not approach zero asymptotically; indeed it diverges, although this is not apparent in the displayed range. But now one observes that  the mass function $m(r)$  grows steeply in the displayed range, whereas the metric function $e^{-\delta(r)}$ appear to converge for large $r$. The solution extends smoothly through both horizons; both $R$ and Kretschmann scalar are finite as $r\to r_h$ and $r\to r_c$.
Indeed, one can check this by obtaining a power series of the solution, valid  close
to the BH/cosmological horizon. But asymptotically, the solutions do not approach de Sitter spacetime.
 
\begin{figure}[t!]
\hbox to\linewidth{\hss%
	\resizebox{9cm}{7cm}{\includegraphics{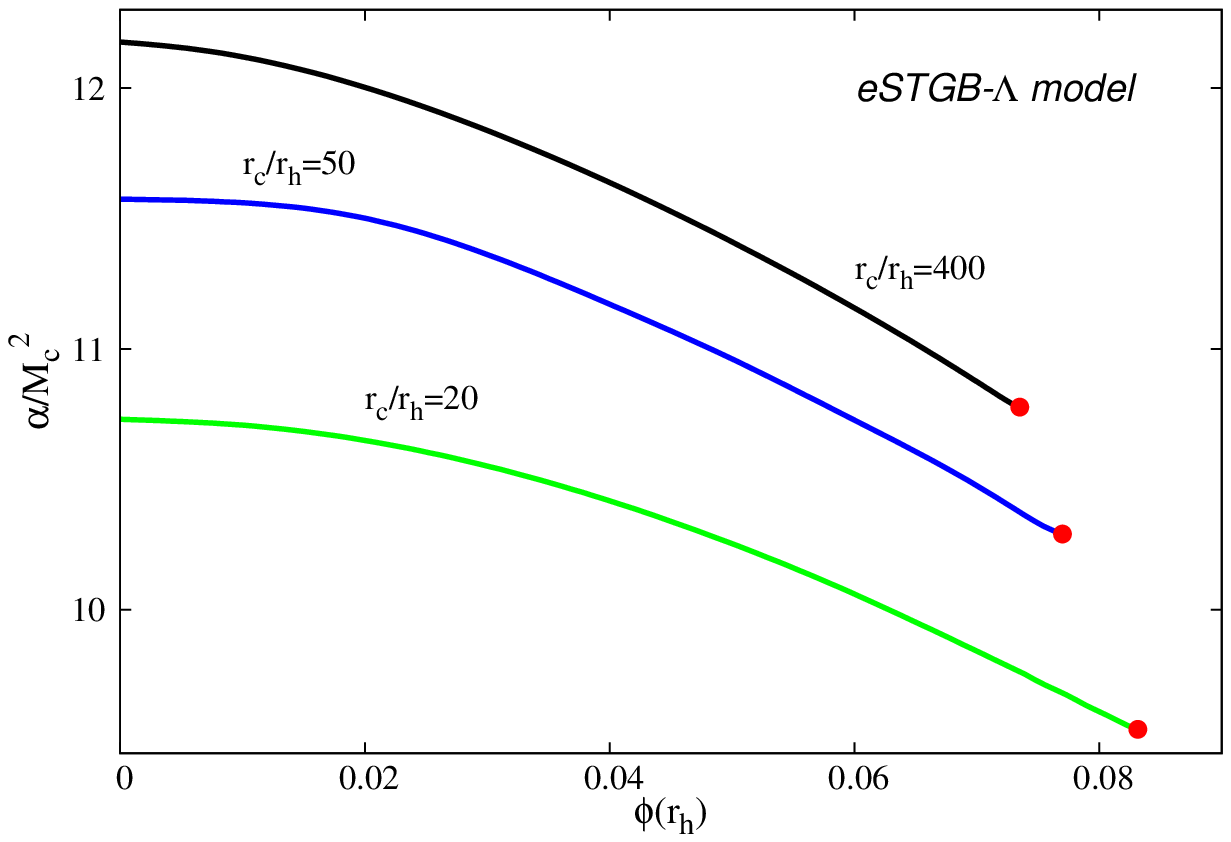}} 
		\resizebox{9cm}{7cm}{\includegraphics{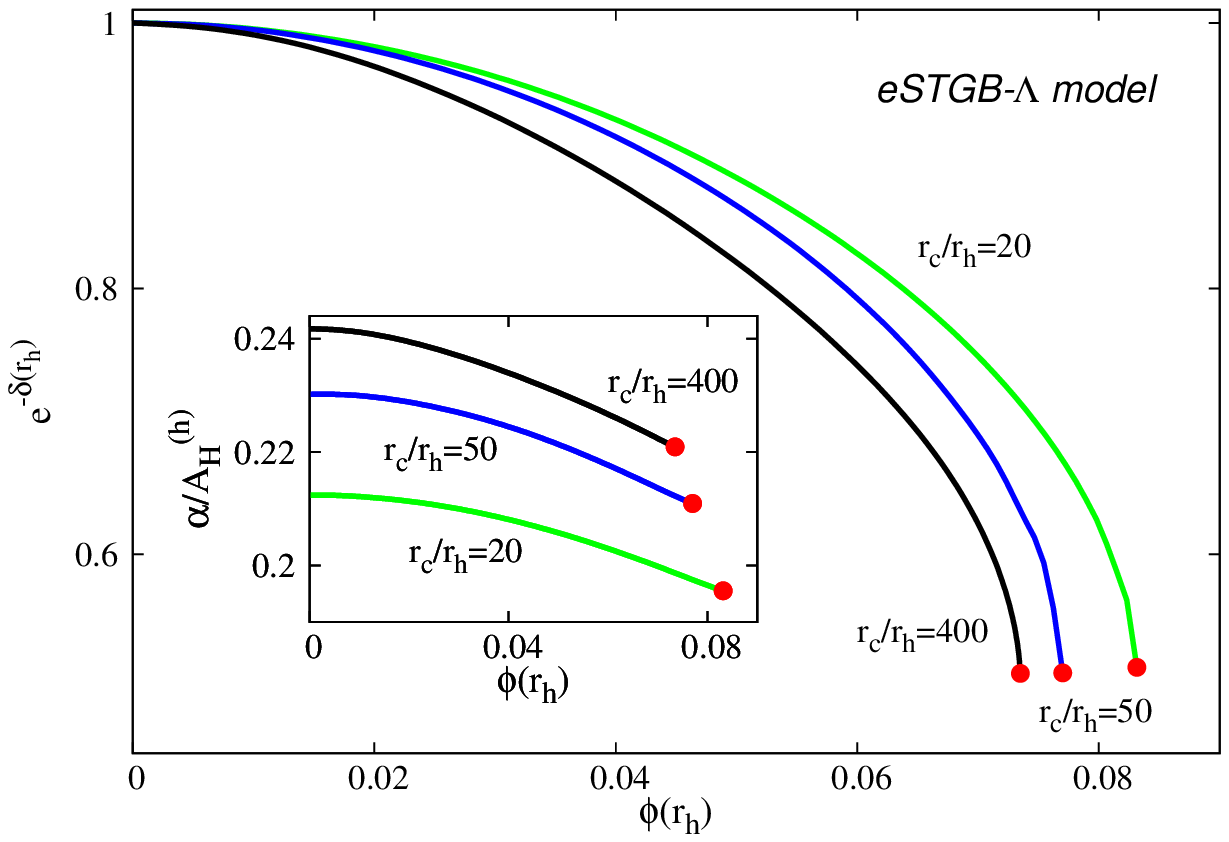}} 

\hss}
\caption{\small
``Mass" (left panel),  BH horizon area and the value of the metric function $e^{-\delta(r)}$ at the horizon (right panel) for eSTGB-$\Lambda$ BHs $vs.$ the scalar field at the BH horizon, for different values of $r_c/r_h$. 
The red dots indicate the critical configurations where the branches stop to exist.
}
\label{EGBs}
\end{figure}

Conveying a more global perspective of the domain of existence of these solutions leads to the following remarks. Similarly to the $\Lambda=0$ case, 
a branch of eSTGB-$\Lambda$ BHs  bifurcates from any  zero mode.
In appropriate variables,  these eSTGB-$\Lambda$ solutions form a line, starting from the smooth $\Lambda$-vacuum limit, as $\phi \to 0$, and ending at a limiting solution - Fig.~\ref{EGBs}.
The existence of this limiting solution can be understood by noticing that,
similarly to the $\Lambda=0$ case~\cite{Silva:2017uqg,Doneva:2017bvd,Antoniou:2017acq},
the nonlinearity associated with the Gauss-Bonnet term 
implies that the derivative of the scalar field at $r=r_h$ solves a second order equation in terms of  $\phi(r_h)$, $\Lambda$ and $\alpha$ (the same holds at the cosmological horizon).
Then $\phi'(r_h)$ becomes imaginary for some critical configuration, and as result
the numerical iterations fail to converge. The ``mass" $M_c=m(r_c)$, BH horizon area and the value of the metric function $e^{-\delta(r)}$ at the BH horizon are shown in  Fig. \ref{EGBs} for the eSTGB-$\Lambda$ BHs as a function of the scalar field at the BH horizon, with $\phi(r_h)=0$ corresponding to the SdS limit. The red dots marking the critical configurations.

As before, we first numerically integrated the field equations between the BH and cosmological horizon.  In a second step, the solutions were extended to the region $r>r_c$.
For all configurations we considered,  the scalar field diverges for  $r \to \infty$, 
a feature inherited from the test field limit. As a result, the mass function diverges as 
$
m(r)\sim r^{ \frac{3}{2}(1+\sqrt{1+16 \alpha \Lambda/ 9})}
$
which implies
$
N(r)\sim r^{ \frac{1}{2}(1+3\sqrt{1+16 \alpha \Lambda/9})}>r^2.
$
This means the solutions do not approach a dS spacetime at future/past infinity. The tachyonic scalar field dominates the behaviour asymptotically. This is (likely) a manifestation of the cosmological instability in eSTGB models discussed in~\cite{Anson:2019uto}.

 {\small \hspace*{3.cm}{\it  } }

 \section{Further remarks}
 \label{section5}
In this work we have studied the impact of a positive cosmological constant on two paradigmatic models of BH spontaneous scalarisation. For $\Lambda=0$, their electrovacuum BH solutions may become
spontaneously scalarised, due to a tachyonic instability triggered by scalar perturbations
\cite{Silva:2017uqg,Doneva:2017bvd,Antoniou:2017acq,Herdeiro:2018wub}.

Our study shows  that the response of the two models, that share many features for $\Lambda=0$, to a non-zero cosmological constant is quite different. While the solutions of the EMS-$\Lambda$ model share the key properties of their asymptotically flat counterparts, with mild differences only, 
the eSTGB-$\Lambda$ model differs from both their flat spacetime counterpart and the EMS-$\Lambda$.
This difference can be traced to the different asymptotic behaviour 
of the source term ${\cal I}$ in the action (\ref{actionS}).
For both models, the scalar field acquires an {\it effective}
tachyonic mass $\mu$ for a region close to the BH horizon.
However, while  for the
EMS-$\Lambda$  the scalar field becomes massless as $r \to \infty$
(the square of the effective field mass
being proportional with Maxwell invariant $F^2$),
this is not the case for the  eSTGB-$\Lambda$ model.
In the latter, $\mu^2$ approaches
asymptotically a  negative value, being proportional to the Gauss-Bonnet invariant for dS spacetime.
As a result, the scalar field diverges in the far field, which results in non-dS asymptotics
of the solutions, despite the presence of a cosmological horizon.
At the same time, the considered configurations 
are regular in the region between the BH and cosmological horizon.

While the results in this work have been found for a quadratic coupling of the scalar field,
we expect that the basic features do not depend  on this specific choice  of the coupling function.
As a direction of further research, it would be interesting to investigate the stability of the EMS-$\Lambda$ solutions.

\section*{Acknowledgements}

This work is supported by the Fundacao para a Ci\^encia e a Tecnologia (FCT) project UID/MAT/04106/2019 (CIDMA), by CENTRA (FCT) strategic project UID/FIS/00099/2013, by national funds (OE), through FCT, I.P., in the scope of the framework contract foreseen in the numbers 4, 5 and 6 of the article 23, of the Decree-Law 57/2016, of August 29,
changed by Law 57/2017, of July 19. We acknowledge support  from the project PTDC/FIS-OUT/28407/2017. This work has further been supported by  the  European  Union's  Horizon  2020  research  and  innovation  (RISE) programmes H2020-MSCA-RISE-2015 Grant No.~StronGrHEP-690904 and H2020-MSCA-RISE-2017 Grant No.~FunFiCO-777740. The authors would like to acknowledge networking support by the
COST Actions CA16104 and CA18108.

 \begin{small}
 
 \end{small}


\begin{thebibliography}{99} 
\bibitem{Perlmutter:1998np}
  S.~Perlmutter {\it et al.}  [Supernova Cosmology Project Collaboration],
  Astrophys.\ J.\  {\bf 517} (1999) 565
  [arXiv:astro-ph/9812133];
\bibitem{Riess:1998cb}
  A.~G.~Riess {\it et al.}  [Supernova Search Team Collaboration],
  Astron.\ J.\  {\bf 116} (1998) 1009
  [arXiv:astro-ph/9805201].
\bibitem{Strominger:2001pn}
  A.~Strominger,
  JHEP {\bf 0110} (2001) 034
  [arXiv:hep-th/0106113];
\%
\bibitem{Witten:2001kn}
  E.~Witten,
  arXiv:hep-th/0106109.
\bibitem{Chrusciel:2012jk}
  P.~T.~Chrusciel, J.~Lopes Costa and M.~Heusler,
  Living Rev.\ Rel.\  {\bf 15} (2012) 7
  [arXiv:1205.6112 [gr-qc]].
\bibitem{Herdeiro:2015waa}
  C.~A.~R.~Herdeiro and E.~Radu,
  Int.\ J.\ Mod.\ Phys.\ D {\bf 24} (2015) no.09,  1542014
  [arXiv:1504.08209 [gr-qc]].
\bibitem{Cai:1997ij}
  R.~G.~Cai, J.~Y.~Ji and K.~S.~Soh,
  Phys.\ Rev.\ D {\bf 58} (1998) 024002
  [gr-qc/9708064].
\bibitem{Torii:1998ir}
  T.~Torii, K.~Maeda and M.~Narita,
  Phys.\ Rev.\ D {\bf 59} (1999) 064027
  [gr-qc/9809036].
\bibitem{Torii:1999uv}
  T.~Torii, K.~Maeda and M.~Narita,
  Phys.\ Rev.\ D {\bf 59} (1999) 104002.
\bibitem{Bhattacharya:2007zzb}
  S.~Bhattacharya and A.~Lahiri,
  Phys.\ Rev.\ Lett.\  {\bf 99} (2007) 201101
  [gr-qc/0702006 [GR-QC]].
\bibitem{Winstanley:2005fu}
  E.~Winstanley,
  Class.\ Quant.\ Grav.\  {\bf 22} (2005) 2233
  [gr-qc/0501096].
\bibitem{Martinez:2002ru}
  C.~Martinez, R.~Troncoso and J.~Zanelli,
  Phys.\ Rev.\ D {\bf 67} (2003) 024008
  [hep-th/0205319].
\bibitem{Schunck:2003kk}
  F.~E.~Schunck and E.~W.~Mielke,
  Class.\ Quant.\ Grav.\  {\bf 20} (2003) R301
  [arXiv:0801.0307 [astro-ph]].
\bibitem{Herdeiro:2014goa}
  C.~A.~R.~Herdeiro and E.~Radu,
  Phys.\ Rev.\ Lett.\  {\bf 112} (2014) 221101
  [arXiv:1403.2757 [gr-qc]].
\bibitem{Brihaye:2005an}
  Y.~Brihaye and T.~Delsate,
  Mod.\ Phys.\ Lett.\ A {\bf 21} (2006) 2043
  [hep-th/0512339].
 
\bibitem{Torii:1995wv}
  T.~Torii, K.~i.~Maeda and T.~Tachizawa,
  Phys.\ Rev.\ D {\bf 52} (1995) R4272
  [gr-qc/9506018].
\bibitem{Breitenlohner:2004fp}
  P.~Breitenlohner, P.~Forgacs and D.~Maison,
  Commun.\ Math.\ Phys.\  {\bf 261} (2006) 569
  [gr-qc/0412067].

\bibitem{Brihaye:2005ft}
  Y.~Brihaye, B.~Hartmann and E.~Radu,
  Phys.\ Rev.\ Lett.\  {\bf 96} (2006) 071101
  [hep-th/0508247].
\bibitem{Brihaye:2006kn}
  Y.~Brihaye, B.~Hartmann, E.~Radu and C.~Stelea,
  Nucl.\ Phys.\ B {\bf 763} (2007) 115
  [gr-qc/0607078].
\bibitem{Silva:2017uqg}
  H.~O.~Silva, J.~Sakstein, L.~Gualtieri, T.~P.~Sotiriou and E.~Berti,
  Phys.\ Rev.\ Lett.\  {\bf 120} (2018) no.13,  131104
  [arXiv:1711.02080 [gr-qc]].
\bibitem{Doneva:2017bvd}
  D.~D.~Doneva and S.~S.~Yazadjiev,
  Phys.\ Rev.\ Lett.\  {\bf 120} (2018) no.13,  131103
  [arXiv:1711.01187 [gr-qc]].
\bibitem{Antoniou:2017acq}
  G.~Antoniou, A.~Bakopoulos and P.~Kanti,
  Phys.\ Rev.\ Lett.\  {\bf 120} (2018) no.13,  131102
  [arXiv:1711.03390 [hep-th]].
\bibitem{Antoniou:2017hxj}
  G.~Antoniou, A.~Bakopoulos and P.~Kanti,
  Phys.\ Rev.\ D {\bf 97} (2018) no.8,  084037
  [arXiv:1711.07431 [hep-th]].
\bibitem{Myung:2018iyq}
  Y.~S.~Myung and D.~C.~Zou,
  Phys.\ Rev.\ D {\bf 98} (2018) no.2,  024030
  [arXiv:1805.05023 [gr-qc]].
\bibitem{Blazquez-Salcedo:2018jnn}
  J.~L.~Blazquez-Salcedo, D.~D.~Doneva, J.~Kunz and S.~S.~Yazadjiev,
  Phys.\ Rev.\ D {\bf 98} (2018) no.8,  084011
  [arXiv:1805.05755 [gr-qc]].
\bibitem{Doneva:2018rou}
  D.~D.~Doneva, S.~Kiorpelidi, P.~G.~Nedkova, E.~Papantonopoulos and S.~S.~Yazadjiev,
  Phys.\ Rev.\ D {\bf 98} (2018) no.10,  104056
  [arXiv:1809.00844 [gr-qc]].
\bibitem{Minamitsuji:2018xde}
  M.~Minamitsuji and T.~Ikeda,
  Phys.\ Rev.\ D {\bf 99} (2019) no.4,  044017
  [arXiv:1812.03551 [gr-qc]].
\bibitem{Silva:2018qhn}
  H.~O.~Silva, C.~F.~B.~Macedo, T.~P.~Sotiriou, L.~Gualtieri, J.~Sakstein and E.~Berti,
  Phys.\ Rev.\ D {\bf 99} (2019) no.6,  064011
  [arXiv:1812.05590 [gr-qc]].
\bibitem{Brihaye:2018grv}
  Y.~Brihaye and L.~Ducobu,
  Phys.\ Lett.\ B {\bf 795} (2019) 135
  [arXiv:1812.07438 [gr-qc]].
\bibitem{Macedo:2019sem}
  C.~F.~B.~Macedo, J.~Sakstein, E.~Berti, L.~Gualtieri, H.~O.~Silva and T.~P.~Sotiriou,
  Phys.\ Rev.\ D {\bf 99} (2019) no.10,  104041
  [arXiv:1903.06784 [gr-qc]].
\bibitem{Doneva:2019vuh}
  D.~D.~Doneva, K.~V.~Staykov and S.~S.~Yazadjiev,
  Phys.\ Rev.\ D {\bf 99} (2019) no.10,  104045
  [arXiv:1903.08119 [gr-qc]].
\bibitem{Myung:2019wvb}
  Y.~S.~Myung and D.~C.~Zou,
  Int.\ J.\ Mod.\ Phys.\ D {\bf 28} (2019) no.09,  1950114
  [arXiv:1903.08312 [gr-qc]].
\bibitem{Andreou:2019ikc}
  N.~Andreou, N.~Franchini, G.~Ventagli and T.~P.~Sotiriou,
  Phys.\ Rev.\ D {\bf 99} (2019) no.12,  124022
  [arXiv:1904.06365 [gr-qc]].
\bibitem{Minamitsuji:2019iwp}
  M.~Minamitsuji and T.~Ikeda,
  Phys.\ Rev.\ D {\bf 99} (2019) no.10,  104069
  [arXiv:1904.06572 [gr-qc]].
\bibitem{Cunha:2019dwb}
  P.~V.~P.~Cunha, C.~A.~R.~Herdeiro and E.~Radu,
  Phys.\ Rev.\ Lett.\  {\bf 123} (2019) no.1,  011101
  [arXiv:1904.09997 [gr-qc]].
\bibitem{Konoplya:2019fpy}
  R.~A.~Konoplya, T.~Pappas and A.~Zhidenko,
  arXiv:1907.10112 [gr-qc].

	

	
\bibitem{Herdeiro:2019yjy}
  C.~A.~R.~Herdeiro and E.~Radu,
  Phys.\ Rev.\ D {\bf 99} (2019) no.8,  084039
  [arXiv:1901.02953 [gr-qc]].
\bibitem{Brihaye:2018bgc}
  Y.~Brihaye, C.~Herdeiro and E.~Radu,
  Phys.\ Lett.\ B {\bf 788} (2019) 295
  [arXiv:1810.09560 [gr-qc]].
\bibitem{Herdeiro:2018wub}
  C.~A.~R.~Herdeiro, E.~Radu, N.~Sanchis-Gual and J.~A.~Font,
  Phys.\ Rev.\ Lett.\  {\bf 121} (2018) no.10,  101102
  [arXiv:1806.05190 [gr-qc]].
\bibitem{Myung:2018vug}
  Y.~S.~Myung and D.~C.~Zou,
  Eur.\ Phys.\ J.\ C {\bf 79} (2019) no.3,  273
  [arXiv:1808.02609 [gr-qc]].
\bibitem{Myung:2018jvi}
  Y.~S.~Myung and D.~C.~Zou,
  Phys.\ Lett.\ B {\bf 790} (2019) 400
  [arXiv:1812.03604 [gr-qc]].
\bibitem{Fernandes:2019rez}
  P.~G.~S.~Fernandes, C.~A.~R.~Herdeiro, A.~M.~Pombo, E.~Radu and N.~Sanchis-Gual,
  Class.\ Quant.\ Grav.\  {\bf 36} (2019) no.13,  134002
  [arXiv:1902.05079 [gr-qc]].
\bibitem{Brihaye:2019kvj}
  Y.~Brihaye and B.~Hartmann,
  Phys.\ Lett.\ B {\bf 792} (2019) 244
  [arXiv:1902.05760 [gr-qc]].
\bibitem{Myung:2019oua}
  Y.~S.~Myung and D.~C.~Zou,
  Eur.\ Phys.\ J.\ C {\bf 79} (2019) no.8,  641
  [arXiv:1904.09864 [gr-qc]].
\bibitem{Konoplya:2019goy}
  R.~A.~Konoplya and A.~Zhidenko,
  Phys.\ Rev.\ D {\bf 100} (2019) no.4,  044015
  [arXiv:1907.05551 [gr-qc]].
\bibitem{Fernandes:2019kmh}
  P.~G.~S.~Fernandes, C.~A.~R.~Herdeiro, A.~M.~Pombo, E.~Radu and N.~Sanchis-Gual,
  arXiv:1908.00037 [gr-qc].
 

 
 
\bibitem{Ramazanoglu:2019gbz}
  F.~M.~Ramazanoglu,
  Phys.\ Rev.\ D {\bf 99} (2019) no.8,  084015
  [arXiv:1901.10009 [gr-qc]].


 
\bibitem{Bakopoulos:2018nui}
  A.~Bakopoulos, G.~Antoniou and P.~Kanti,
  Phys.\ Rev.\ D {\bf 99} (2019) no.6,  064003
  [arXiv:1812.06941 [hep-th]].
\bibitem{Balasubramanian:2001nb}
  V.~Balasubramanian, J.~de Boer and D.~Minic,
  Phys.\ Rev.\ D {\bf 65} (2002) 123508
  doi:10.1103/PhysRevD.65.123508
  [hep-th/0110108].
	
\bibitem{COLSYS}
 U. Ascher, J. Christiansen, R.~D. Russell,
 Math. Comp. {\bf 33} (1979) 659;
\\
 U. Ascher, J. Christiansen, R.~D. Russell,
 ACM Trans. {\bf 7} (1981) 209.	

\bibitem{Brihaye:2016vkv}
  Y.~Brihaye, C.~Herdeiro and E.~Radu,
  Phys.\ Lett.\ B {\bf 760} (2016) 279
  [arXiv:1605.08901 [gr-qc]].
\bibitem{Astefanesei:2003gw}
  D.~Astefanesei, R.~B.~Mann and E.~Radu,
  JHEP {\bf 0401} (2004) 029
  [hep-th/0310273].


\bibitem{Brill:1993tw}
  D.~R.~Brill and S.~A.~Hayward,
  Class.\ Quant.\ Grav.\  {\bf 11} (1994) 359
  [gr-qc/9304007].

\bibitem{Romans:1991nq}
  L.~J.~Romans,
  Nucl.\ Phys.\ B {\bf 383} (1992) 395
  [hep-th/9203018].

 \bibitem{Nariai}
H.~Nariai,
Sci. Rep. Tohoku Univ. Eighth Ser. {\bf 34} (1950) 160 .

\bibitem{Hod:2019pmb}
  S.~Hod,
  Phys.\ Rev.\ D {\bf 100} (2019) no.6,  064039.
	
\bibitem{Anson:2019uto}
  T.~Anson, E.~Babichev, C.~Charmousis and S.~Ramazanov,
  JCAP {\bf 1906} (2019) 023
  [arXiv:1903.02399 [gr-qc]].


 \end{thebibliography}
 \end{document}